\documentstyle[preprint,aps,epsfig]{revtex}
\begin{document}

\tighten
\draft
\preprint{
\vbox{
\hbox{June 1998}
\hbox{FZ-IKP(TH)-1998-10}
\hbox{ADP-98-16/T292}
}}

\title{Semi-inclusive Pion Production and the $d/u$ Ratio}
\author{W. Melnitchouk}
\address{Institut fuer Kernphysik,
	Forschungszentrum Juelich,
	D-52425 Juelich, Germany	\\
	{\tt w.melnitchouk@fz-juelich.de}}
\author{J. Speth}
\address{Institut fuer Kernphysik,
	Forschungszentrum Juelich,
	D-52425 Juelich, Germany	\\
	{\tt j.speth@fz-juelich.de}}
\author{A.W. Thomas}
\address{Department of Physics and Mathematical Physics,
	and Special Research Center for the
	Subatomic Structure of Matter,
	University of Adelaide,
	Adelaide 5005, Australia	\\
	{\tt athomas@physics.adelaide.edu.au}}

\maketitle

\begin{abstract}
We explore the feasibility of directly extracting the large-$x$ valence
$d/u$ ratio through a measurement of pions in the current fragmentation
region of semi-inclusive deep-inelastic scattering from protons.
\end{abstract}


\vspace*{1cm}

The $d/u$ ratio contains important information about the flavor
structure of the proton.
In particular, its asymptotic behavior at large $x$ can tell us
which mechanism is responsible for the breaking of
SU(2)$_{\rm spin} \times$ SU(2)$_{\rm flavor}$ symmetry.
Given that there are firm predictions for this behavior,
it constitutes a serious test of perturbative QCD \cite{FJ}.

So far, a direct measurement of the $d/u$ ratio has been rather
difficult, mainly because the cross sections decrease rapidly in
the extreme kinematics near $x \sim 1$.
Previous analyses \cite{SLAC,EMC} have used inclusive deep-inelastic
scattering (DIS) data on proton and deuteron targets to obtain the
neutron to proton structure function ratio, from which $d/u$ can be
extracted at large $x$ according to:
\begin{eqnarray}
\label{F2np}
{ F_2^n \over F_2^p }
&=& { 1 + 4 d/u \over 4 + d/u },\ \ \ \ \ x \rightarrow 1.
\end{eqnarray}

However, the neutron structure function in Eq.(\ref{F2np}) is
obtained from data on the deuteron, which suffer from the fact
that nuclear effects, even in the deuteron, become quite significant
at large $x$ \cite{SMEAR,OTHER}.
In particular, whether one corrects for Fermi motion only, or in
addition for binding and nucleon off-shell effects, the extracted
neutron structure functions for $x \agt 0.7$ can differ rather
dramatically \cite{NP}.

The question is therefore how to avoid the problem of the uncertainty
in the extraction procedure introduced by nuclear effects.
A number of suggestions how to overcome this problem have been
raised in the literature.
Perhaps the simplest idea is to use neutrino and antineutrino beams
to scatter on proton targets, thereby selecting the $d$ and $u$
quark flavors, respectively.
Although such data exist \cite{NEUTRINO}, their statistical accuracy is
unfortunately not very high. Furthermore,
the range of $x$ covered ($x \alt 0.6$) is too
small to constrain the asymptotic behavior.
Another recent suggestion was to use $W$-boson production in $pp$
and $p\bar p$ collisions \cite{W}, measuring the decay muon asymmetry
at large lepton rapidities.
While simple in principle, this method requires one to be
at the current limit of rapidity space, and to make stringent cuts
on the transverse energy of the muons produced,
which reduces the statistical accuracy considerably.

An alternative approach is to use semi-inclusive DIS, in which a hadron
is tagged in the final state in coincidence with the scattered lepton.
With a deuteron target, one could in principle minimize the nucleon
off-shell ambiguities by observing a recoil proton in the target
fragmentation region, which would be indicative of the primary
scattering having taken place from a near on-shell neutron \cite{SEMID}.
Such experiments are indeed already being discussed in connection
with Jefferson Lab and HERMES, although one must keep in mind the
potential problem of the $Q^2$ not being high enough at these facilities
to enable a clean separation of the target and current regions.

A related idea is to use semi-inclusive DIS data on hadron production
in the current fragmentation region to measure the relative yields of
$\pi^+$ and $\pi^-$ mesons.
In fact, the HERMES Collaboration has extracted $d/u$ from the
$\pi^+ - \pi^-$ difference, on both protons and bound neutrons
\cite{HERMES}.
The idea here is fairly simple: at large $z$ the $u$ quark fragments
primarily into a $\pi^+$, while a $d$ fragments into a $\pi^-$, so that
at large $x$ and $z$ one could have a direct measure of the valence
$d/u$ quark ratio.
Note that one should not be too close to $z=1$, as in that region
the fragmentation process may no longer be incoherent and may not be
factorizable into a partonic cross section and a target-independent
fragmentation function.

To describe the semi-inclusive process, one requires the fragmentation
functions of quarks into pions to be known.
Fortunately, the fragmentation functions for the leading
$u \rightarrow \pi^+$ and non-leading $u \rightarrow \pi^-$ processes
have been measured by the European Muon Collaboration at CERN
\cite{EMCFRAG}, albeit under the assumption of charge symmetry,
$D_u^{\pi^+} = D_d^{\pi^-}$,
and that the fragmentation is target-independent.

The apparent advantage of using both $p$ and $n$ targets at HERMES
is that all dependence on the fragmentation functions cancels, thereby
removing any uncertainty that might be introduced by poor knowledge of
the hadronization process.
On the other hand, the disadvantage of this approach is that one is
still left with the same problem of using the deuteron as an effective
neutron target, together with the inherent difficulties in unfolding
the neutron structure function, which, as we argue below, are likely
to be more problematic than the uncertainty in the fragmentation
functions.
The main contribution that such an approach could make would
be to serve as a check of the extraction of the $d/u$ ratio
from existing inclusive DIS data on $p$ and $D$.
It would be preferable, given the original aim of avoiding nuclear
uncertainties, to use only $p$ targets, together with the empirical
information on the $z$-dependence of the fragmentation functions.

In the parton model the number of pions in a given $x$ and $z$ bin
can be written as a product of a quark distribution function, $q^h(x)$,
in the hadron $h$, and a fragmentation function giving the probability
of the scattered quark $q$ producing a pion:
\begin{eqnarray}
N_h^\pi &\sim& \sum_q\ e_q^2\ q^h(x)\ D_q^\pi(z).
\end{eqnarray}
(The scale dependence of both the parton distribution and fragmentation
functions is suppressed.)
For a proton target, therefore, one has:
\begin{mathletters}
\begin{eqnarray}
N_p^{\pi^+} &\sim& 4 u(x)\ D(z)\ +\ d(x)\ \bar D(z),	\\
N_p^{\pi^-} &\sim& 4 u(x)\ \bar D(z)\ +\ d(x)\ D(z),
\end{eqnarray}
\end{mathletters}%
where $D(z) \equiv D_u^{\pi^+} = D_d^{\pi^-}$ is the
leading fragmentation function (assuming isospin symmetry),
and $\bar D(z) \equiv D_d^{\pi^+} = D_u^{\pi^-}$ is the
non-leading fragmentation function.

In addition to protons, the HERMES Collaboration also uses
deuteron targets.
Using isospin symmetry for the parton distributions, $u^p = d^n$,
etc., and assuming that the deuteron is a system of two bound nucleons,
one can write for the deuteron:
\begin{mathletters}
\begin{eqnarray}
N_D^{\pi^+} &\sim& (\tilde u(x) + \tilde d(x))\ (4 D(z) + \bar D(z)),	\\
N_D^{\pi^-} &\sim& (\tilde u(x) + \tilde d(x))\ (4 \bar D(z) + D(z)),
\end{eqnarray}
\end{mathletters}%
where $\tilde q$ denotes the smeared quark distribution of a nucleon
bound in the deuteron, which can be approximated by:
\begin{eqnarray}
\tilde q(x) &=& \int {dy \over y} f_{N/D}(y)\ q(x/y).
\end{eqnarray}
The smearing function, $f_{N/D}(y)$, gives the probability (in the
infinite momentum frame) of finding a nucleon in the deuteron with
light-cone momentum fraction $y$, and is given explicitly in
Ref.\cite{SMEAR}, for example.
At small $x$ the difference between $q$ and $\tilde q$ is less than
$\sim 2\%$, although at large $x$, where the $d/u$ ratio is less well
known, the difference can be much larger.

Combining the proton and deuteron cross sections, one can
define the bound neutron ($\tilde n$) distribution,
$N_{\tilde n}^\pi \equiv N_D^{\pi} - N_p^{\pi}$:
\begin{mathletters}
\begin{eqnarray}
N_{\tilde n}^{\pi^+}
&\sim& 4 (\tilde d(x) + \epsilon_u(x))\ D(z)\
       + (\tilde u(x) + \epsilon_d(x))\ \bar D(z),	\\
N_{\tilde n}^{\pi^-}
&\sim& 4 (\tilde d(x) + \epsilon_u(x))\ \bar D(z)\
       + (\tilde u(x) + \epsilon_u(x))\ D(z),
\end{eqnarray}
\end{mathletters}%
where $\epsilon_q(x) \equiv \tilde q(x) - q(x)$.
Taking the difference between the $\pi^+$ and $\pi^-$ cross
sections for the proton and bound neutron gives:
\begin{mathletters}
\begin{eqnarray}
N_p^{\pi^+} - N_p^{\pi^-}
&\sim& (4 u(x) - d(x)) (D(z) - \bar D(z)),		\\
N_{\tilde n}^{\pi^+} - N_{\tilde n}^{\pi^-}
&\sim&	\left( 4 \tilde d(x) - \tilde u(x)
	     + 4 \epsilon_u(x) - \epsilon_d(x)
	\right) (D(z) - \bar D(z)).
\end{eqnarray}
\end{mathletters}%
The ratio of these,
\begin{eqnarray}
\label{Rnp}
R_{np}(x,z)\ \equiv\
{ N_{\tilde n}^{\pi^+} - N_{\tilde n}^{\pi^-}
 \over N_p^{\pi^+} - N_p^{\pi^-} }
&=& { 4 \tilde d(x)  - \tilde u(x)
    + 4 \epsilon_u(x) - \epsilon_d(x)
\over 4 u(x) - d(x) },
\end{eqnarray}
is then independent of the fragmentation function,
and is a function of $x$ only.

Clearly, under the assumption that nuclear corrections
to the quark distributions are negligible, $\tilde q = q$, one has:
\begin{eqnarray}
R_{np} &=& { 4 d(x)/u(x) - 1 \over 4  - d(x)/u(x) }\ .
\end{eqnarray}

In Fig.~1 we show the ratio $R_{np}$ calculated with the assumption of
no nuclear effects (dashed), and with the smearing correction (solid)
which accounts for the Fermi motion and binding effects in the deuteron,
as described in Ref.\cite{SMEAR}.
The parton distributions were taken from the recent CTEQ4
parameterization \cite{CTEQ4} (throughout we use NLO distributions
evaluated in the DIS scheme) at $Q^2=10$ GeV$^2$ in Fig.~1(a),
while in Fig.~1(b) the $d$ quark distribution was modified to
have the correct perturbative QCD limit \cite{FJ}, according
to Refs.\cite{NP,W}:
\begin{eqnarray}
\label{dmod}
{d(x) \over u(x)}
&\rightarrow& \left. {d(x) \over u(x)}\right|_{\rm CTEQ4} + \Delta(x).
\end{eqnarray}
The analysis of Ref.\cite{NP} suggested that at $Q^2 \approx 10$ GeV$^2$
the correction term could be parameterized by the simple form \cite{W}:
\begin{eqnarray}
\label{Delta10}
\Delta(x)
&\approx& 0.2\ x^2\ \exp(-(1-x)^2),\ \ \ \ \ Q^2 \approx 10\ {\rm GeV}^2,
\end{eqnarray}
so that in the limit $x \rightarrow 1$, the modified $d/u \rightarrow 1/5$,
consistent with the expectation from perturbative QCD \cite{FJ}.
Beyond $x \sim 0.7$ the difference between the corrected and
uncorrected ratios is clearly quite dramatic.
Consequently a $d/u$ ratio obtained from such a measurement in
this kinematic region, without correcting for nuclear effects,
would give rather misleading results.

As mentioned above, one can avoid the problem of nuclear corrections
altogether by comparing data for $\pi^+$ and $\pi^-$ production on
proton targets alone.
Although the statistical accuracy will not be as good as when both
protons and deuterons are used, one can expect them to be sufficient
to obtain reliable information on the ratio.
Taking the ratio of the $\pi^-$ to $\pi^+$ proton cross sections,
one finds:
\begin{eqnarray}
\label{Rpi}
R^{\pi}(x,z)\ \equiv\
{ N_p^{\pi^-} \over N_p^{\pi^+} }
&=& { 4 \bar D(z)/D(z) + d(x)/u(x)
\over 4 + d(x)/u(x) \cdot \bar D(z)/D(z) }.
\end{eqnarray}
The fragmentation functions $D$ and $\bar D$, measured by the EMC
\cite{EMCFRAG}, are shown in Fig.~2, together with parameterizations
for the leading and non-leading functions:
\begin{mathletters}
\label{Dz}
\begin{eqnarray}
D(z) &=& 0.7\ (1-z)^{1.75} / z,		\\
\bar D(z) &=& { (1-z) \over (1+z) } D(z).
\end{eqnarray}
\end{mathletters}%
In the limit $z \rightarrow 1$, the leading fragmentation function
clearly dominates, $D(z) \gg \bar D(z)$.
In this case the ratio $R^{\pi} \rightarrow (1/4) d/u$ \cite{LAND}.
Although the point $z=1$ cannot be reached experimentally, for
reference we show in Fig.~3 the theoretical ratio expected in this
limit for two different parameterizations, the standard CTEQ4
distributions \cite{CTEQ4}, and with the $d$ quark distribution
modified according to Eq.(\ref{dmod}), both at $Q^2=10$ GeV$^2$.
The $z=1$ limit provides the maximal difference between the two ratios.

In the realistic case of smaller $z$, the $\bar D/D$ term in
Eq.(\ref{Rpi}) will contaminate the yield of fast pions originating
from struck primary quarks, diluting the cross section with pions
produced from secondary fragmentation picking up $q\bar q$ pairs
from the vacuum.
Nevertheless, one can estimate the yields of pions using the
empirical fragmentation functions in Eq.(\ref{Dz}).
Integrating the differential cross section over a range of $z$,
as is more practical experimentally, the resulting ratios for cuts
of $z > 0.3$ and $z > 0.5$ are shown in Fig.~4.
One sees that decreasing the lower limit for $z$ has the effect of
raising the cross section ratio significantly, because of the larger
integrated contribution from the non-leading fragmentation, which
is more important at smaller $z$.
Although the relative difference between the ratios for the two
forms of asymptotic $d/u$ behavior then becomes smaller, the absolute
difference between these remains relatively constant, and should be
measurable with the high luminosities available at current facilities.

For some final words of caution, we should note that the assumptions
of factorization and neglect of higher twists are more questionable
in the very large $x$ region.
As one approaches the elastic limit at $x=1$, the effective value
of $Q^2$ where a leading twist treatment is likely to be sufficient
will increase with $x$ like $Q^2/(1-x)$.
{}From inclusive DIS data, where scaling sets in already at
$Q^2 \sim 2$ GeV$^2$, one expects a leading twist analysis
for $Q^2 = 10$ GeV$^2$ to be valid up to $x \sim 0.8$.
For semi-inclusive DIS, however, it is not known up to which value
of $x$ the leading twist perturbative QCD framework can be applicable
at any given $Q^2$, and ultimately this question can only be answered
experimentally.
One may expect that the $Q^2$ effects would not be as important for
the ratios of cross sections discussed in this paper as for the
absolute cross sections.
To this effect we have also calculated $R^\pi$ at $Q^2 = 100$ GeV$^2$,
and indeed find the differences with the results in Fig.4 to be rather
small.

In summary, we reiterate the importance of an accurate experimental
determination of the behavior of the valence $d/u$ ratio as
$x \rightarrow 1$.
Not only are the present fits to world data in clear disagreement
with the predictions of perturbative QCD (unlike the reanalysis of
Ref.\cite{NP}), but the discrepancy is extremely important when it
comes to estimating event rates for charged current events at
the large values of $x$ and $Q^2$ accessible at HERA \cite{CHARM}.
Our analysis of binding and Fermi motion corrections in the deuteron
suggests that it is not profitable to try to resolve this issue by
combining semi-inclusive data on protons and deuterons.
On the other hand, the issue could be decided through semi-inclusive
pion production measurements on the proton alone.
While this will require a dedicated experiment in the kinematically
less favored regime where the energy transfer to the final pion
is larger than $z \sim 0.5$, in view of the importance of settling
this question we believe it is crucial that the effort be made soon.

\acknowledgements

We would like to thank B. Kopeliovich and N. Isgur for useful
discussions.
W.M. and J.S. would like to thank the Special Research Centre for
the Subatomic Structure of Matter at the University of Adelaide
for support during a recent visit, where part of this work was
performed.
This work was supported by the Australian Research Council.


\begin{figure}
\label{fig1}
\epsfig{figure=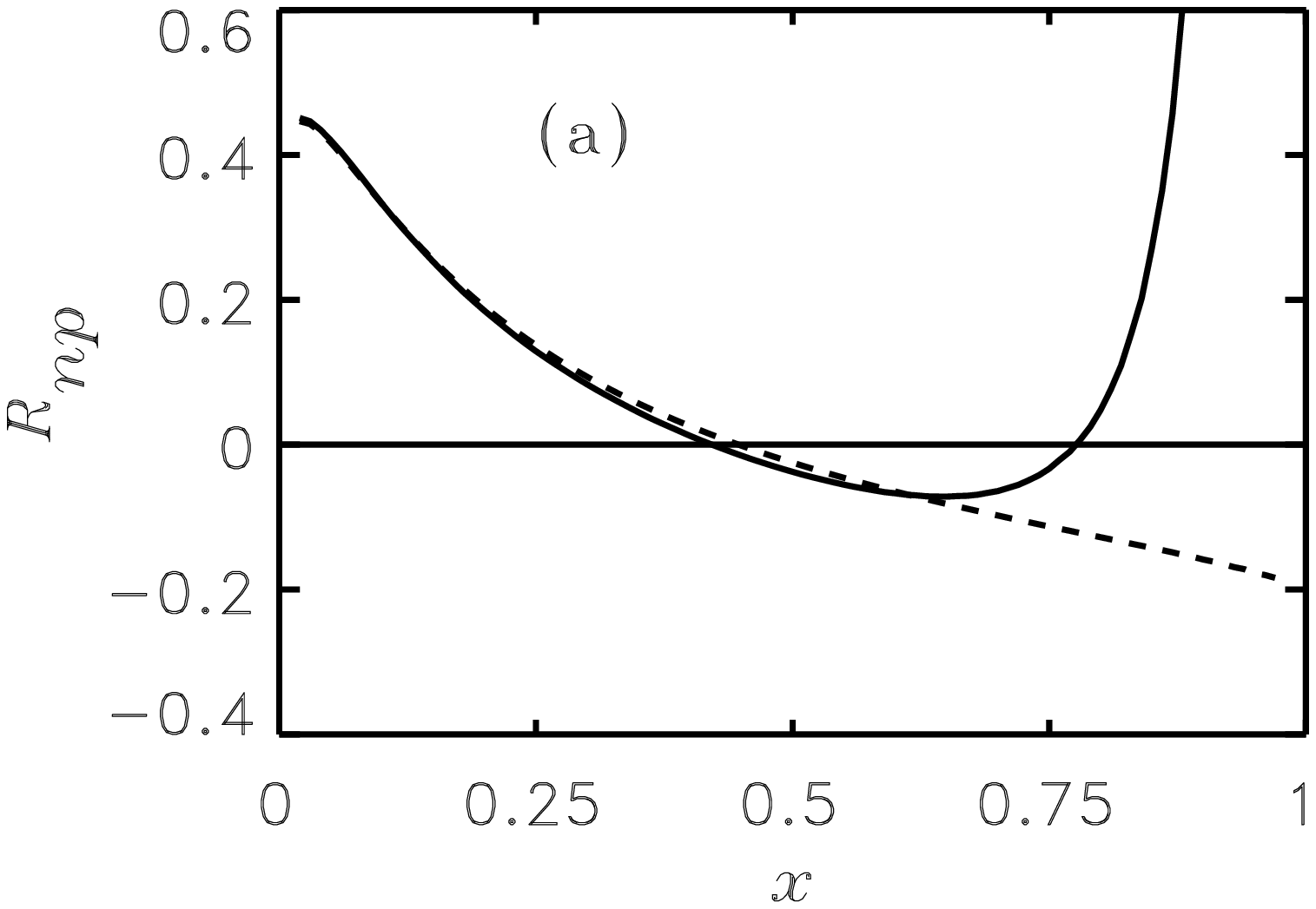,height=8.5cm}
\epsfig{figure=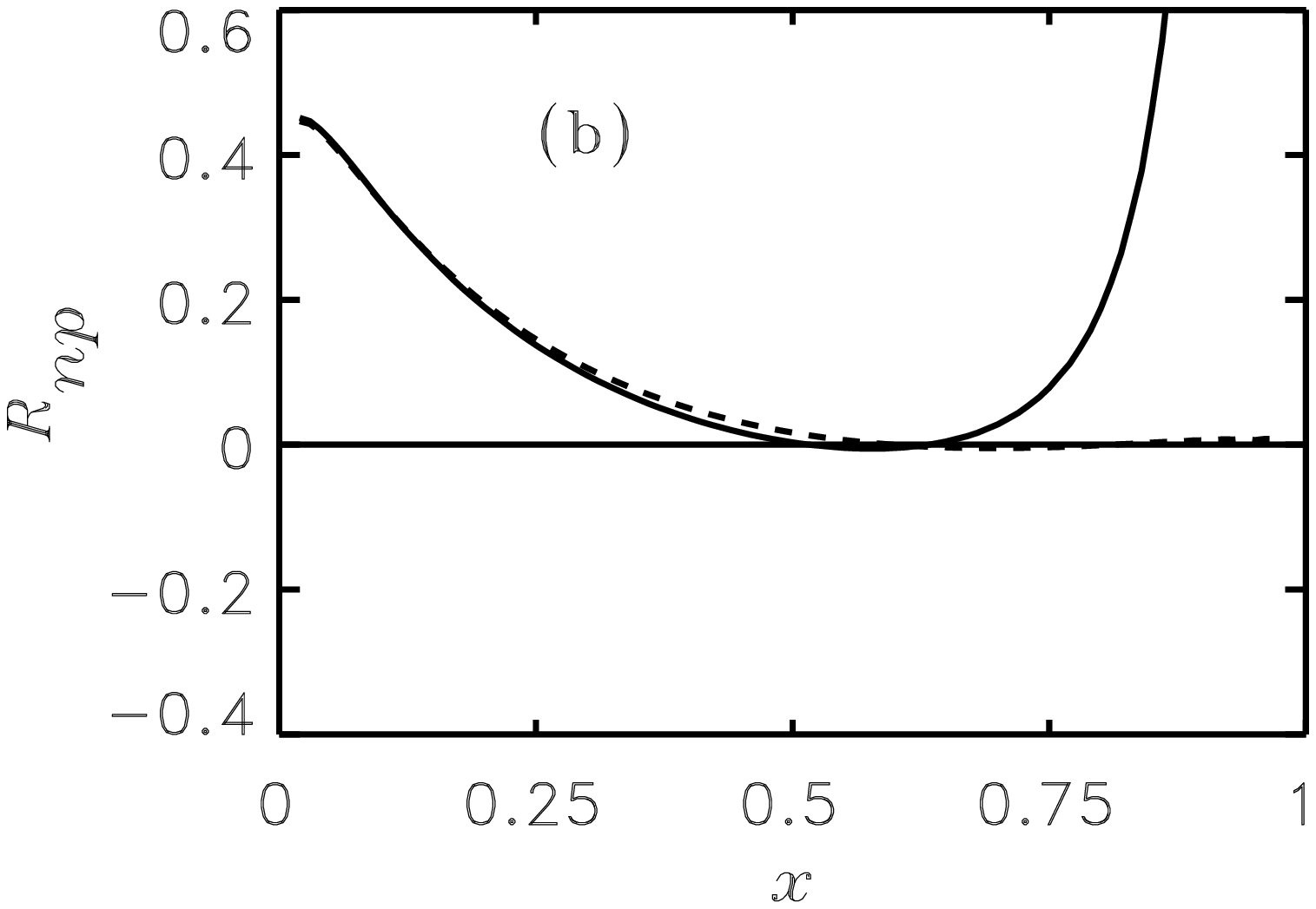,height=8.5cm}
\caption{Ratio $R_{np}$ in Eq.(\protect\ref{Rnp}) calculated with (solid)
	and without (dashed) smearing corrections.  The $u$ and $d$
	distributions were taken (a) from the CTEQ4 parameterization
	\protect\cite{CTEQ4}, and (b) with the $d$ quark distribution
	modified as in Ref.\protect\cite{W} to have the correct
	perturbative QCD limit \protect\cite{FJ,NP}.}
\end{figure}

\begin{figure}
\label{fig2}
\epsfig{figure=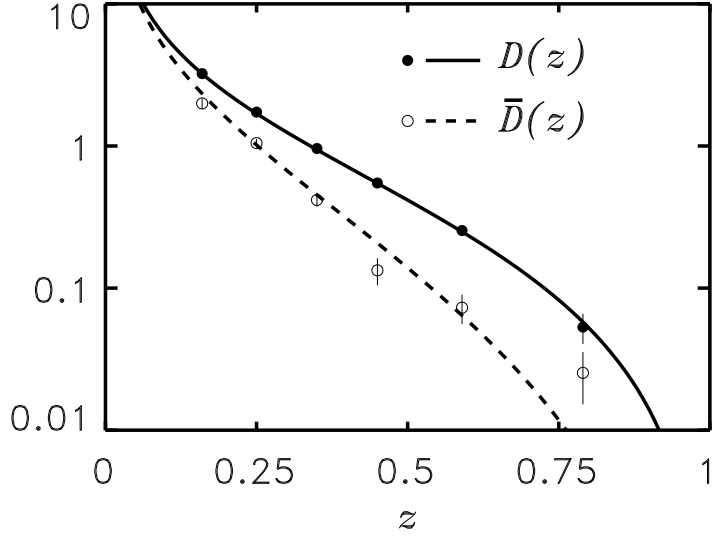,height=8.5cm}
\caption{Fragmentation functions, $D(z)$ and $\bar D(z)$, from the EMC
	experiment \protect\cite{EMCFRAG}, together with the parameterizations
	given in Eq.(\protect\ref{Dz}).}
\end{figure}

\begin{figure}
\label{fig3}
\epsfig{figure=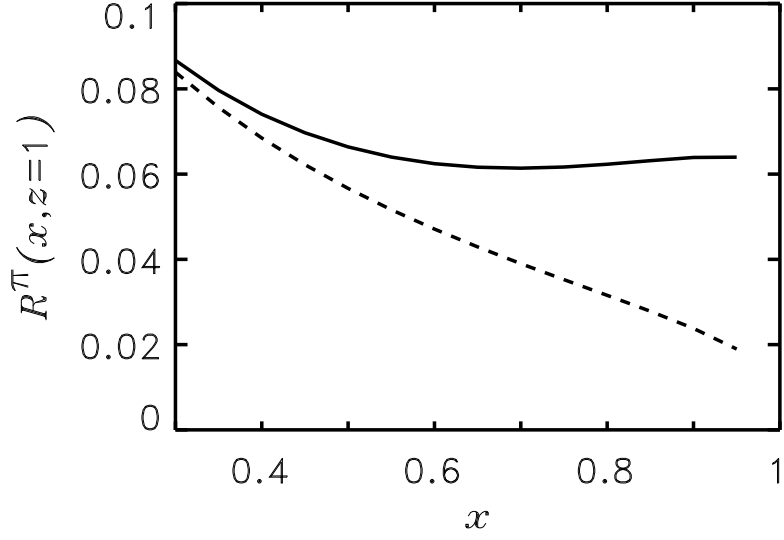,height=8.5cm}
\caption{Theoretical ratio, $R^{\pi}$, as a function of $x$ for
	fixed $z=1$.  The dashed line represents the ratio constructed
	from the CTEQ4 parameterization \protect\cite{CTEQ4}, while
	the solid includes the modified $d$ distribution according
	to Eq.(\protect\ref{dmod}).}
\end{figure}

\begin{figure}
\label{fig4}
\epsfig{figure=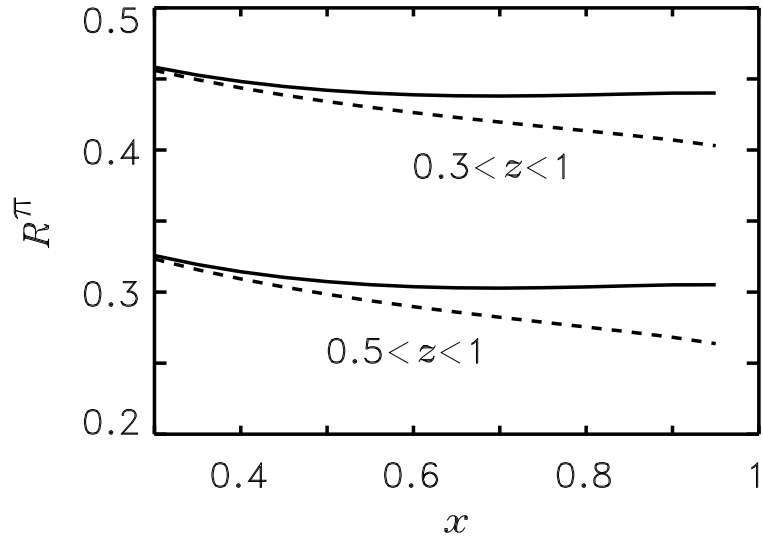,height=8.5cm}
\caption{Ratio $R^{\pi}$ as function of $x$, integrated over $z$
	between $0.3 < z < 1$ and $0.5 < z < 1$.  The solid and
	dashed curves are as in Fig.3.}
\end{figure}

\end{document}